\documentclass[aps,letterpaper,english,reprint,showkeys,floatfix]{revtex4-1}
\usepackage[T1]{fontenc}
\usepackage[latin9]{inputenc}
\usepackage{babel}
\usepackage{calc}
\usepackage{amsmath}
\usepackage{amssymb}
\usepackage{graphicx}
\usepackage{wrapfig}
\usepackage{float}
\usepackage{cleveref}
\usepackage{multirow}
\usepackage[left]{lineno} % default option is 'left'

\usepackage{setspace}
\usepackage{verbatim}

\setcitestyle{authoryear}
\makeatletter
\usepackage{parskip} 

\pdfpageheight\paperheight
\pdfpagewidth\paperwidth

\providecommand{\LyX}{\texorpdfstring%
  {L\kern-.1667em\lower.25em\hbox{Y}\kern-.125emX\@}
  {LyX}}
\DeclareMathOperator{\argminH}{argmin}   % Jan Hlavacek

\makeatother

\begin{document}

\title{A probabilistic atlas for cell identification}
\author{Greg Bubnis}
\author{Steven Ban}
\author{Matthew D. DiFranco}
\author{Saul Kato}
\email{corresponding author: saul.kato@ucsf.edu}

\selectlanguage{english}%

\affiliation{University of California, San Francisco~\\
 Weill Institute for Neurosciences, Department of Neurology }

\date{\today}

\begin{abstract}

\noindent We propose a general framework for a collaborative machine learning system to assist bioscience researchers with the task of labeling specific cell identities from microscopic still or video imaging. The distinguishing features of this approach versus prior approaches include: (1) use of a statistical model of cell features that is iteratively improved, (2) generation of probabilistic guesses at cell ID rather than single best-guesses for each cell, (3) tracking of joint probabilities of features within and across cells, and (4) ability to exploit multi-modal features, such as cell position, morphology, reporter intensities, and activity. We provide an example implementation of such a system applicable to labeling fluorescently tagged \textit{C. elegans} neurons. As a proof of concept, we use a generative spring-mass model to simulate sequences of cell imaging datasets with variable cell positions and fluorescence intensities. Training on synthetic data, we find that atlases that track inter-cell positional correlations give higher labeling accuracies than those that treat cell positions independently. Tracking an additional feature type, fluorescence intensity, boosts accuracy relative to a position-only atlas, suggesting that multiple cell features could be leveraged to improve automated label predictions.
\end{abstract}

\keywords{cell atlas, Bayesian, cell identification, cell type, calcium imaging}
\maketitle

%============================================================================================

\section{\label{sec:level1}Introduction}

The development of effective fluorescent reporters, increase in computing power, and proliferation of volumetric microscopy techniques are now enabling acquisition of large tissue volumes or entire organisms, within which many individual cells can be discerned (\Citealt{Ahrens:2013, Kato:2015, Prevedel:2014, Skocek:2018}). With the explosion in data collection capabilities and throughput, the time demand for the experimentalist is shifting from data acquisition to post-acquisition data analysis. 

Common post-acquisition analyses are to (1) find and delineate these cells in space and, for video recordings, in time, often collectively referred to as region-of-interest (ROI) detection, segmentation, and tracking, (2) extract static or time-series data of fluorescence intensity arising from calcium dynamics or other cellular processes, and (3) determine the identity or type of the delineated cells, in order to fuse cellular-resolution data across trials and animals (``inter-trial registration'') and pave the way for statistically powerful analysis. This last step is the focus of this paper.

In eutelic organisms or certain experimental contexts in non-eutelic organisms, cell identity labels may be unique within a recording (``cell ID''); in other contexts, the labels may be non-unique (``cell type''). Here, we focus on the former problem, typified by volumetric recordings of fluorescently tagged neurons of the brain of the nematode \textit{C. elegans} hermaphrodite, which is composed of precisely 302 neurons. Manual identification based on comparison to previously labeled images or a traditional reference atlas is extremely laborious, and there currently does not exist a widely used method for automated cell identification in this context. We propose a robust and practical approach for automated cell ID.
%============================================================================================
\section{\label{sec:level1}Design Features}
We outline what we believe to be necessary design requirements as follows:

\emph{Statistical model.} A canonical reference atlas of anatomy is a timeworn pedagogical device, but well appreciated to neglect inherent biological variability. A more informative atlas \textendash \ derived from experimental data across multiple trials \textendash \  should maintain a record of the biological variability of anatomical or physiological features (\citealt{ProbAtlas1,ProbAtlas2,ProbAtlas3}). Specifically, the atlas should maintain a probability distribution for each atlas feature or, to capture relationships between these features, a joint probability distribution over the set of atlas features.

\emph{Incrementally trained with partially labeled data.} We would like the quality of our atlas to improve as each new training dataset is contributed to the system, as measured by a declining error rate and increasing confidence levels of labeling. Furthermore, training sets are likely to lack complete labeling, possibly severely so, because manual identification of even a subset of neurons is extremely laborious and unreliable.

\begin{figure}[h]
    \centering
    \includegraphics[width = 0.47\textwidth]{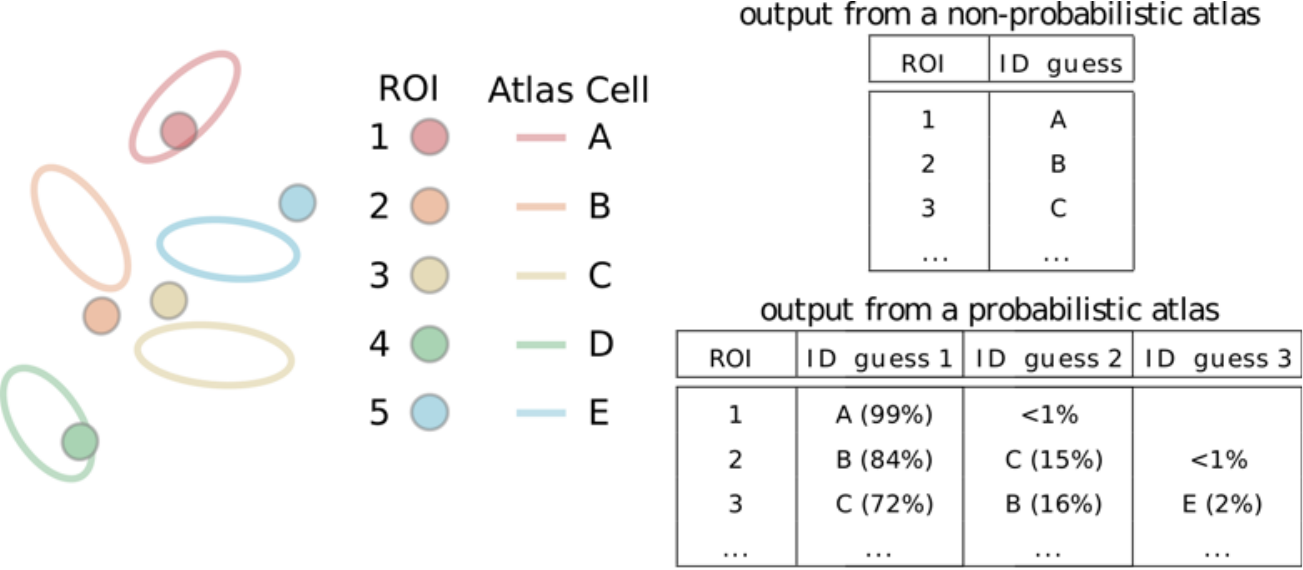}
    \caption{Example probabilistic and non-probabilistic output for a system containing 5 unlabeled ROIs and atlas cells. In this case, labeling is only performed on $x,y$ positions of ROI centroids (dots), and the atlas contains information on the biological variability of the cells (ellipses). }
    \label{fig:probatlas}
\end{figure}

\emph{Probabilistic output.}  Earlier approaches to produce algorithms for cell identification, such as those based on bipartite graph matching, were prone to cascading errors where the mislabeling of one cell could lead to the mislabeling of many other cells, prompting the exploration of probabilistic approaches (\citealt{Myers:2009, Myers:2014, Wu:ensemble}). It is also probably unreasonable to expect a perfect automatic labeling for any large dataset due to the noise and variability in all imaging of biological systems, and lack of perfect knowledge in real-world experimental scenarios. 
Therefore, a system which can report label guesses with marginal probabilities, which reflect confidence of labeling, is of more practical use (Figure \ref{fig:probatlas}).

\emph{Acceptance of datasets with missing or erroneous data.} Variability in development, reporter expression patterns, feature extraction algorithms, and imaging contexts may result in incoming datasets having varying numbers of cells to be identified. To be generally applicable, identification should be possible on ROI sets that correspond to both subsets and supersets of atlas cells.

\emph{Flexibility in exploiting available cell properties.} Cells have many observable features which may covary within and between cells, including position, size, shape, fluorescence intensity, and time series (Table \ref{table:features}). We want the atlas to leverage this covariance structure to ID cells with increased confidence.

\emph{Community driven.} Each experimental context may require its own atlas depending on the characteristics of the data; however, there are likely to be several groups around the world working in similar contexts who need a cell ID system and who can, in return, contribute training datasets for the improvement of the atlas. The precise parameters of each experiment are likely to change, and the use of new datasets must be robust to these changes.

We recommend that this system be implemented as a cloud service for accessibility and to pool the efforts of researchers.

\begin{table}[ht]
\caption{Potential features of cells to learn}
\begin{tabular}{l|l|l|l|}
\cline{2-4}
 & \textbf{Data Type}    & \textbf{Data Format} & \textbf{Dimension}         \\ \cline{2-4} 
\multirow{7}{*}{\rotatebox[origin=c]{90}{$~~~~$Single Frame}} 
& Position   & $x,y,z$     & 3   \\
& Size       & $\sigma_x,\sigma_y,\sigma_z$ & 3\\
& Orientation  & $\phi, \theta, \psi$  & 3       \\
& Fluorescence intensity  & $i_{\text{GFP}}, i_{\text{RFP}}$,... &   $[1-5]$ \\
& Expression pattern   & fold(gene)  &  $N_{\text{genes}}$     \\
& Morphology           & voxel set  & $X\times  Y\times Z$      \\
\cline{2-4}
\multirow{2}{*}{\rotatebox[origin=c]{90}{Video}} & Fluorescence time series           & $i_{1,..T}(t)$         & $[1-5]\times T$           \\
& Spacetime trajectory & $\mathbf{s}_{1,..T}(t)$    & $3\times T$                          \\ \cline{2-4} 
\end{tabular}
\label{table:features}
\end{table}

%==================================================================================
\section{\label{sec:level2}System Description}
\begin{figure}[H]
\centering
\includegraphics[scale=0.75]{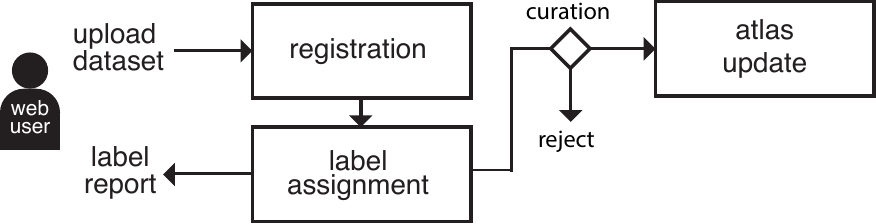}
\caption{Atlas system flowchart}
\label{fig:flow}
\end{figure}

We now describe a system that satisfies the aforementioned design requirements.

A sequence of datasets, possibly contributed by different users, is processed in the cloud one at a time by the system (Figure \ref{fig:flow}). We assume that a dataset arrives post-ROI detection (and post time series extraction for a video) as a set of unlabeled feature records, one for each cell. 
For ease of reference, visualization, and proof checking by end users, datasets may be accompanied by the original raw image stack data, although the atlas does not make use of this raw data.

To enable effective atlas learning, datasets must first be registered to match the coordinate system of the reference atlas without the benefit of labels to match particular cells from dataset to particular cells in the atlas. This registration operation may involve global transformations of the dataset, such as scaling, translation, and rotation, as well  more localized transformations such as volumetric warping to correct for tissue deformation. Registration of other feature types, such as fluorescence intensity, may involve other approaches such as histogram equalization or color space transformation. 

After registration, labels must be assigned to each cell in the dataset. Label assignment across all cells in a dataset is a combinatorial optimization problem, in which an objective function representing likelihood is maximized. Estimated marginal probabilities of individual cell labelings are reported to the user (Figure \ref{fig:probatlas}), along with accompanying visualization data for ease of validation.

A \textit{curation} step, either manual or automated, follows: if the quality the dataset and success of labeling is acceptable, then the labeled dataset, possibly in part or in whole, will be used for a training update to the atlas. In the early stages of atlas training, manual curation will likely be required to produce a useful system for early users. In the future, partial or weighted-importance acceptance of datasets could replace this binary acceptance procedure.

%==================================================================================
\section{Mathematical Setting}

We follow a sequential Bayesian updating approach. We define a \textit{probabilistic atlas} as a set of hyperparameters $\phi$ that describe a joint probability distribution of cell parameters $\theta$, along with the selection of probability distribution function family $f$:
\begin{equation}
p(\theta_{1...N}|\phi) = f(\phi)
\end{equation}
with $N$ being the number of atlas cells.

Each dataset $\mathbf{x}$ comes into the system as an unrolled, concatenated feature vector ($\mathbf{x}_1,\mathbf{x}_2,..,\mathbf{x}_j,..,\mathbf{x}_M$) where $\mathbf{x}_j$ contains features for ROI (i.e. unlabeled cell) $j$. At present we only consider the case of $M=N$. For each $\mathbf{x}$, global registration (spatial and/or in other feature spaces) is performed between dataset and atlas features, which produces a transformed dataset $\hat{\mathbf{x}}$:
\begin{equation}
\hat{\mathbf{x}}=T(\mathbf{x},\hat{\theta}_T)   
\label{eq:basic_atlas}
\end{equation}
where parameter vector $\hat{\theta}_T$ is the result of optimizing over an objective function such as least-squares error to nearest neighbor. Without such global registration, direct application of probability models with location parameters is likely to fail. An example algorithm for spatial registration is \textit{coherent point drift} (CPD) (\citealt{Myronenko:hc}). We exclude these transformation parameters from our atlas hyperparameters as they are likely to be dependent on the nature of particular datasets, though in the future a supervised learning approach could be applied to the registration task.

After spatial registration, a putative labeling $L$, defined as a mapping from atlas cells $i\in \{1,2,3,.. N\}$ to ROIs $j \in \{1,2,3,.. N\}$ with $L_i=j$ indicating that atlas cell $i$ is matched to ROI $j$, is determined by maximizing likelihood over possible label permutations. This is a combinatorial search problem, and our strategies to solve it are discussed later in section \ref{algo}.  

For a transformed, labeled dataset that passes curation, denoted by $L^*(\hat{\mathbf{x}})$, a sequential Bayesian update of the hyperparameters is performed:
\begin{equation}
\phi_{new}=g(\phi_{old},L^*(\hat{\mathbf{x}}))
\end{equation}
where the form of $g$ is determined by solving the Bayes' theorem equation
\begin{equation}
\label{eq:posterior}
p_{conj}(\theta|\phi_{new})=p_{conj}(\theta|\phi_{old},L^*(\mathbf{\hat{x}})) = \frac{p(L^*(\hat{\mathbf{x}})|\theta) p_{conj}(\theta|\phi_{old})}{p(L^*(\hat{\mathbf{x}})|\phi_{old})}
\end{equation}
under the appropriate choice of conjugate prior form $p_{conj}$, which is in turn based on the choice of distribution family $f$ (\citealt{Murphy:2012}). The initial values of hyperparameters may be drawn from other data sources or based on uninformative or reference priors.

We consider three types of probabilistic atlas of increasing expressivity: univariate (UV), cell multivariate (CMV), and full multivariate (FMV). For ease of illustration, we use the normal distribution family, although it is not a requirement of the atlas.

\emph{Univariate model (UV)}. In this case, we assume that the probability distributions of each cell parameter are independent and identically distributed (iid). Each is tracked with a normal distribution, $\mathcal{N}(\mu, \sigma^{2})$ with unknown mean $\mu$ and unknown variance $\sigma^{2}$. Estimates of the true $\mu$ and $\sigma^{2}$ can be obtained from sampled data using the normal-inverse-gamma (NIG) conjugate prior with hyperparameters $(m, V, \alpha, \beta)$ where $m$ and $\beta$ are priors for the mean and variance (up to a scaling factor); and $V$ and $\alpha$ are scalars representing the strength of belief of the respective priors.

\emph{{Cell multivariate model (CMV)}}. In addition to tracking variability of individual features, we may wish to capture relationships of different feature types within each cell, such as correlation between the $x$, $y$, and $z$ coordinates of the cell. We still ignore inter-cell correlation and assume that the cell models are iid. In this case, the probability distribution for each cell is $\mathcal{N}(\mathbf{\mu}, \Sigma)$, where $\mathbf{\mu}$ is a vector of means for each parameter, and $\Sigma$ is the covariance matrix of all features of a single cell, both unknown. 

Estimates of $\mathbf{\mu}$ and $\Sigma$ can be obtained from sample data using the normal-inverse-Wishart (NIW) conjugate prior with hyperparameters $(m_0, S_0, \kappa_0, \nu_0)$ 
where $m_0$ and $S_0$ are priors for the mean and covariance (up to a scaling factor); and $\kappa_0$ and $\nu_0$ are scalars representing the strength of belief of the respective priors.

\emph{{Full multivariate model (FMV)}}. In the full multivariate case, we drop the assumption of iid cells in order to capture inter-cell, inter-feature correlations and therefore use a single multivariate model spanning all parameters. Using a normal distribution, we can employ the same NIW conjugate prior across the full feature set.

In all three cases, we update the atlas hyper parameters with new labeled datasets according to Equation \ref{eq:posterior}. Our atlases then atlas models yield estimates of a multivariate normal distribution, $p(\hat{\mathbf{x}}) = \mathcal{N}(\hat{\mathbf{x}}|\mu,\Sigma)$, but with different learned covariance matrix structure: 
$\Sigma^{\text{UV}}$ is diagonal, $\Sigma^{\text{CMV}}$ is block diagonal with each block modeling intra-cell feature correlations, and $\Sigma^{\text{FMV}}$ has no \textit{a priori} block diagonal structure.

\emph{{Labeling.}} A putative labeling is scored using a cost function defined by the negative log likelihood as a function of atlas parameters:
\begin{equation}
\label{eq:costGeneral}    
\text{cost}(L(\hat{\mathbf{x}})) = -\log p(L(\hat{\mathbf{x}})|\phi)
\end{equation}
For our models, $p(\hat{\mathbf{x}})$ is given by the multivariate Student-T distribution (\citealt{Murphy:2012}). Any convenient cost function that is a monotonic function of the negative log likelihood is also suitable, such as 
\begin{equation}
\label{eq:cost}    
\text{cost}(L(\hat{\mathbf{x}})) = -\log p(L(\hat{\mathbf{x}})|\theta^{\text{MAP}})
\end{equation}
where $\theta^{\text{MAP}}$ are the maximum a posteriori parameters from the atlas posterior (\ref{eq:posterior}).

\textit{Probabilistic output.} The model output would ideally be the marginal probabilities of ROI $j$ having cell label $i$ evaluated over all $N!$ possible labelings:
\begin{equation}
p_{ij} = Z^{-1}\sum_{k=1}^{N!} p(L^{k}(\hat{\mathbf{x}})) \delta(L^{k}_{i}-j)
\label{eq:marginalized}
\end{equation}
where $Z=\sum_{k=1}^{N!} p(L^{k}(\hat{\mathbf{x}}))$, and $p(L(\hat{\mathbf{x}}))$ is the marginal distribution, i.e. integrated over $\theta$. The factorial scaling, however, makes this full evaluation intractable for more than a small number of cells. We presume, without proof, that for a well-fitting model, the terms in the summation for $p_{ij}$ are dominated by a comparatively small number $n\ll N!$ of low-cost (high probability) labelings such that a partial summation
\begin{equation}
\hat{p}_{ij} = \hat{Z}^{-1}\sum_{k=1}^{n} p(L^{k}(\hat{\mathbf{x}})) \delta(L^{k}_{i}-j)
\label{eq:marginalized_truncated}
\end{equation}
where $\hat{Z}$ is analogous to the normalization $Z$ above, is a reasonable estimate to (\ref{eq:marginalized}). Note that in the $n$=1 limiting case, with a single best-guess labeling $L$, we recover the usual non-probabilistic labeling $p_{ij} = \delta(L_{i}-j)$. In our probabilistic scheme, the cost global minimum (GM) $L^{\text{GM}}$ has the most statistical weight in (\ref{eq:marginalized}) and (\ref{eq:marginalized_truncated}), and it is therefore essential to have an algorithm capable of finding it. However, in addition, we seek to sample other low-cost labelings such that the partial summation (\ref{eq:marginalized_truncated}) can be used to approximate the fully marginalized probabilities (\ref{eq:marginalized}).

%==================================================================================
\section{Proof-of-concept atlas: spring-mass model}

\begin{figure}[!htb] 
\includegraphics[width=0.95\columnwidth]{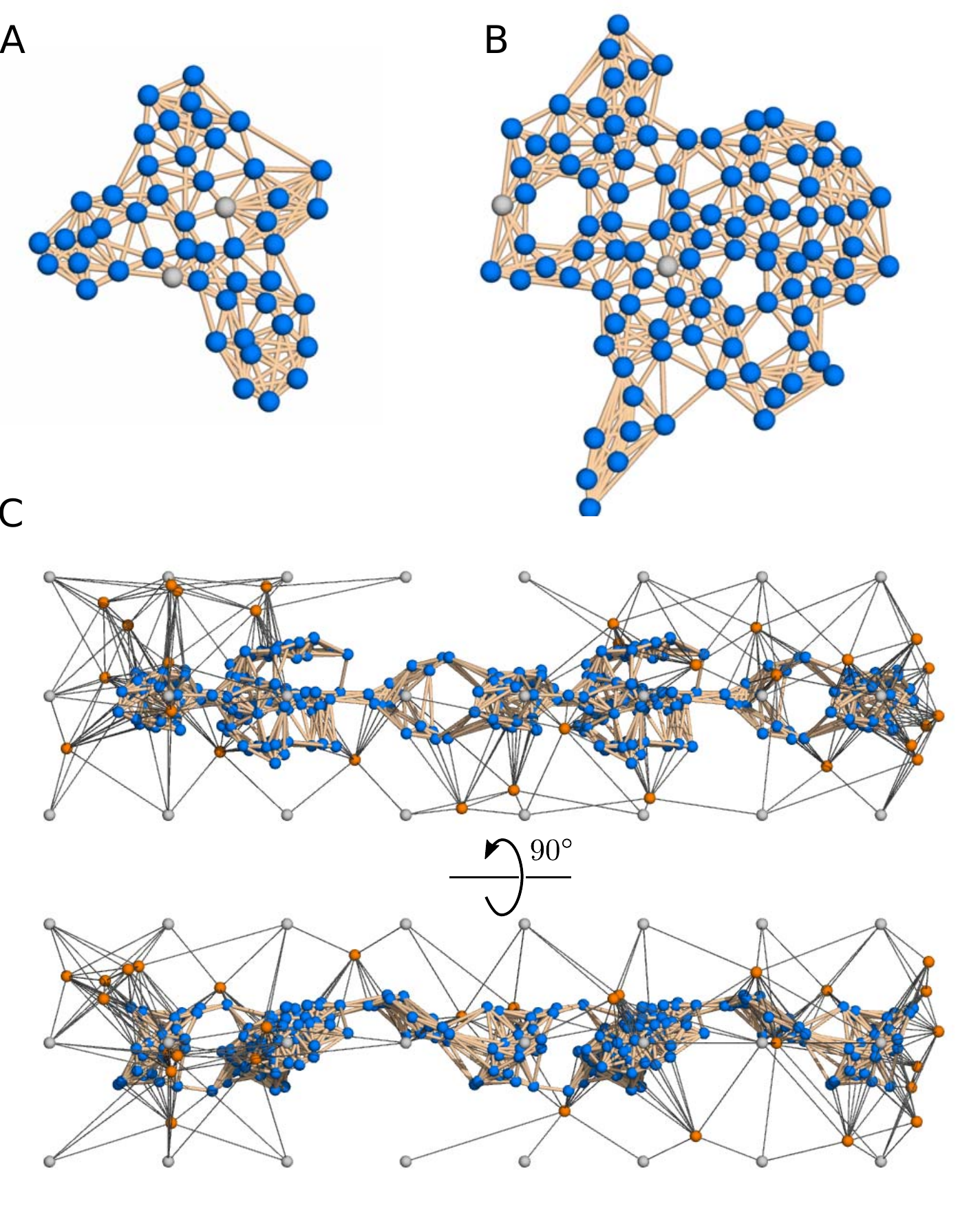}
\caption{\small{Toy models with (A) 50, (B) 100, and (C) 302 cells. Systems (A) and (B) comprise cells (blue) and two stationary anchor points (gray) which are also included in labeling. The $N=302$ system (C) comprises cells (blue spheres) used for labeling, as well as matrix points (orange spheres) and anchor points (gray spheres). In (C), thin lines distinguish cell-matrix or matrix-anchor connections from (thick) cell-cell connections.}}
\label{fig:fig03}
\end{figure}

We illustrate the construction and use of a probabilistic atlas for synthetic data. We generated a sequence of frames by simulating the dynamics of a spring-mass system, with each frame representing a new incoming dataset. This model is intended to capture the spatial variability of cells across experiments. The physical behavior of the toy model loosely models the phenomenon of non-overlapping cells jostling within a confined volume with correlated motion due to deformations of the surrounding tissue. The cell feature set is limited to $x$, $y$, and $z$ positions. In this model, the ground-truth labeling ($L^{\text{GT}}$) is known and the system's behavior is tunable, providing a convenient testbed. Note that in this proof-of-concept atlas, we assume each training frame is accurately labeled, and contains the same number of ROIs as atlas cells.

The generative toy model comprises point masses in 2D or 3D space, representing cell centroids, connected by damped springs. To test different aspects of the atlas labeling scheme, we generated toy systems with 10, 50, 100, and 302 cells with varying positions and connectivity. The cells were embedded in different environments via spring connections to flexible ``matrix'' points and fixed ``anchor'' points.

The $N=10$ system is 2D and sufficiently small such that all $10! \approx 3.6\times10^6$ labelings can be enumerated, thereby allowing comparison between the approximate (\ref{eq:marginalized_truncated}) and exact (\ref{eq:marginalized}) marginal probability outputs. The cells' coordinates were drawn from uniform distributions that allow for significant overlap to test both atlas quality and probabilistic outputs.

The $N=50$ and $N=100$ system (Figure \ref{fig:fig03}A,B and Supplementary Movies 1-2) are 2D and exhibit large-amplitude correlated shifts in cells' positions. The initial coordinates were generated by a branched growth scheme to give a lumpy but compact distribution of points in space.

The $N=302$ system (Figure \ref{fig:fig03}C and Supplementary Movie 3) is based partially on the empirical neuroanatomy of \textit{C. elegans}. A set of 134 head neuron 3D positions derived from (\citealt{White:1986kn}) were copied and translated on the $z$-axis (corresponding to the worm head-tail axis) to give a system of 302 cells with high aspect ratio. This system also incorporates regularly spaced anchor points as well as randomly positioned matrix particles which are bound to both the cells and the anchors, mimicking the influence of surrounding tissue.

The spring-mass systems were simulated via Langevin dynamics (\citealt{schlick2010MD}, see Methods). The particles' $x$, $y$, and $z$ coordinates were snapshotted at fixed time intervals to reduce correlations between time-adjacent frames, and an initial segment of each trajectory was discarded to allow for equilibration of the dynamical behavior. Independent simulations, with different random seeds, were used for atlas training and labeling validation.

\begin{figure*}[!hbt] 
\includegraphics[width=\textwidth]{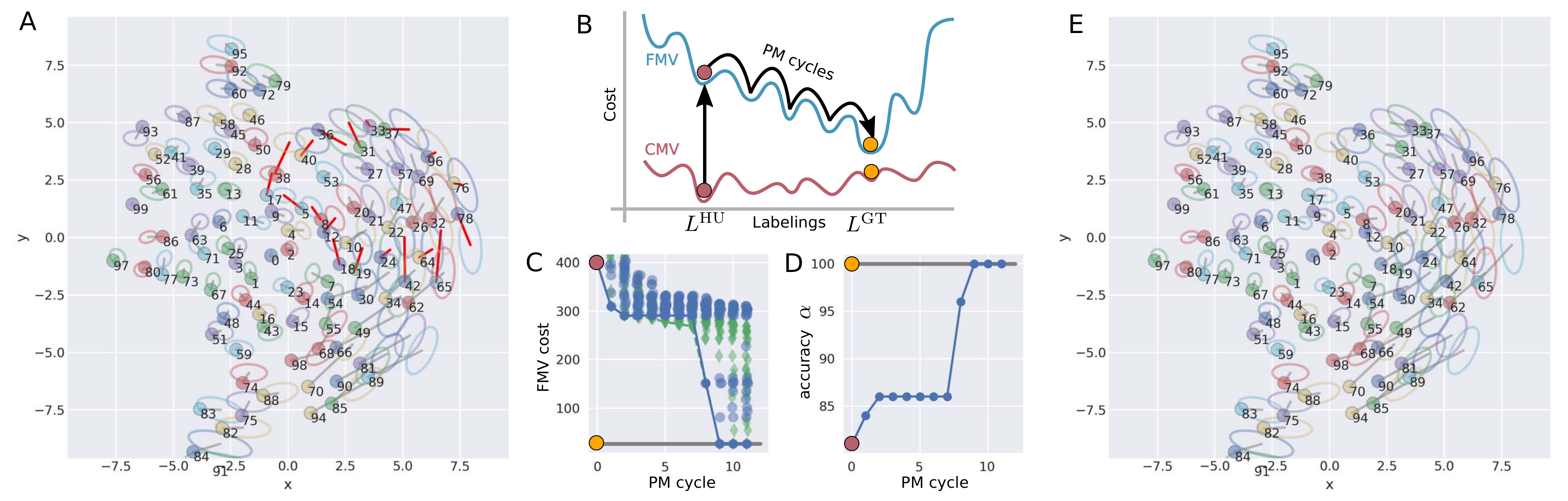}
\caption{\small{ Cascading assignment errors using a CMV atlas and the Hungarian algorithm can be remedied using a FMV atlas and the PM scheme. (A) $L^{\text{HU}}$ obtained using a CMV atlas for the $N=100$ system. Ellipses represent the mean position and covariance ($1\sigma$ contours) for each cell of the CMV atlas and dots are cells to be labeled. Gray/red lines indicate correct/incorrect assignments, respectively. (B) Schematic view of the CMV/FMV cost landscapes and PM search scheme. The HU labeling $L^{\text{HU}}$ (red dot) is the CMV (red curve) cost GM $L^{\text{GM}}$, but not $L^{\text{GT}}$ (orange dot). Starting with $L^{\text{HU}}$, the PM algorithm seeks to find $L^{\text{GT}}$ by minimizing the FMV cost function (blue curve). Note that the vertical cost offset between FMV and CMV is arbitrary and the labeling cost landscapes are, in reality, discrete rather than continuous. (C) FMV labeling costs during PM, with semi-transparent blue circles and green diamonds representing the population pool of legal and illegal assignments, respectively. (D) Labeling accuracy (100 being perfect) during PM for the lowest cost legal labeling. In (C) and (D) the solid blue line and solid circles indicate the lowest cost legal assignment within the population. The accuracy of the lowest cost labeling in (D) is not guaranteed to increase monotonically. (E) The GT assignment determined using PM, where ellipses and dots are as in (A).}}
\label{fig:fig04}
\end{figure*}

%============================================================================================
\section{Labeling algorithm} \label{algo}

To perform atlas guided cell labeling we used the cost function (\ref{eq:cost}), derived from the atlas (UV, CMV, FMV) and an algorithm to search the combinatorial space of labelings. Depending on the atlas model, different schemes are available to solve the combinatorial labeling problem. For the UV and CMV atlas models, the probability factorizes and the cost function becomes a sum of single cell assignment costs,
\begin{equation}
  \begin{aligned}
    \text{cost}(L(\hat{\mathbf{x}})) &= \sum_{i=1}^{N} -\log p_i(\hat{\mathbf{x}}_j|\theta_i^{\text{MAP}})\\
        &= \sum_{i=1}^{N} c_i(\hat{\mathbf{x}}_j) 
  \end{aligned}
\label{eq:costSum}
\end{equation}
In these cases, the task to find $L^{\text{GM}}$ equates to the well-known linear assignment problem (LAP). Deterministic LAP solvers, namely the Hungarian algorithm (HU) (\citealt{Kuhn:1955}), can reach the optimal solution in polynomial time. This straightforward scheme, however, does not guarantee that the result $L^{\text{HU}}$ is the correct labeling $L^{\text{GT}}$, only that it is the optimum, \emph{given the cost function}. Figure \ref{fig:fig04} illustrates how $L^{\text{HU}}$ (for a CMV atlas) can contain assignment errors that `cascade' throughout the structure. 

These issues may be addressable with a full multivariate model (FMV) that captures correlations of features between cells. Unfortunately, the cost function for the FMV model does not factorize; thus the assignment problem is nonlinear. To approach this combinatorial optimization problem (GM search), we use a population cost minimization (PM) scheme that uses greedy and stochastic minimization steps as well as genetic algorithm-like mating operations. Non-bipartite (`illegal') assignments -- where one ROI $j$ may hold two or more atlas cell labels, i.e. $L_{i}=L_{i'}=j$, and other ROIs are unlabeled -- can be generated as intermediates in this procedure, but they are `legalized' by resolving missing and duplicate labels.

\begin{figure}[!hbt]
\centering
\includegraphics[width=0.7 \columnwidth]{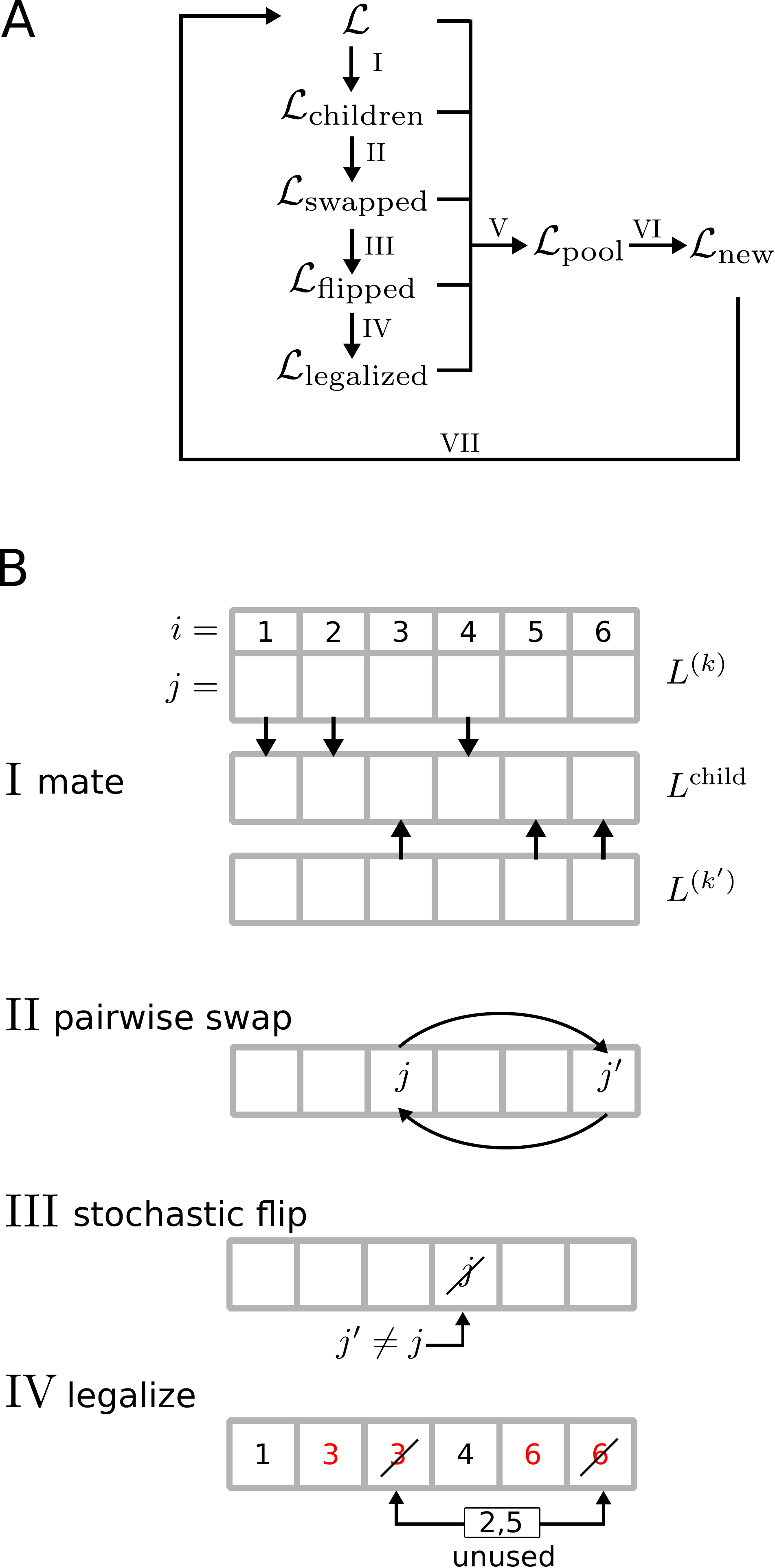}
\caption{\small{(A) Schematic of the PM algorithm. (B) Different moves used during PM. A labeling $L^{(k)}$, shown for the mate move, is illustrated as two stacked rows where each column, an $(i,j)$ pair, indicates a single label assignment $L_i=j$. The upper row of $i$ values is fixed, and is thus omitted for the rest of the illustrated labelings.}}
\label{fig:figMoves}
\end{figure}

The PM scheme is carried out in cycles, where one cycle takes a population of labelings $\mathcal{L}=\{L^{(1)},L^{(2)},L^{(3)},..\}$, performs intermediate steps, and returns a new population $\mathcal{L}_{\text{new}}$. Figure \ref{fig:figMoves} illustrates one cycle (panel A) and its constituent labeling moves (panel B). Each cycle starts (step I) by creating a set of 32 new labelings, $\mathcal{L}_{\text{children}}$, either by mating pairs or copying an individual from $\mathcal{L}$ (with mating probability 50\%). The labelings in $\mathcal{L}_{\text{children}}$ then go through steps II-IV in series before the results are pooled (step V) and culled (step VI). The different intermediate steps/moves are summarized as follows:
\begin{enumerate}
    \item[I. ]Mating: A child labeling $L^{\text{child}}$ is created where each element is drawn with a 50:50 chance from either $L^{(k)}$ or $L^{(k')}$. This introduces randomization and can potentially combine correct labels from disparate parent labelings, but can also create illegal assignments.
    \item[II. ]Pairwise swaps (greedy): For each atlas index $i$ (order shuffled) of labeling $L$, the pairwise swap $L_i\rightarrow L_{i'}, L_{i'}\rightarrow L_i$ where $i'\in\{1,2,3,...N\}$ that gives the lowest cost (including the null swap) is selected. This step tests all pairwise swaps, requiring $\mathcal{O}(N^2)$ cost function calls, assuming no heuristics are applied to winnow attempted swaps.
    \item[III. ]Stochastic flips: One label is flipped, $L_i\rightarrow j' \ne j$, where $i$ and $j'$ are both randomly drawn and the flip is only accepted only if cost decreases (1000 attempts per labeling).
    \item[IV. ] Legalization: For a labeling that is `illegal' due to mating or stochastic flips, for each $i$ (shuffled order) if $L_i=L_{i'}=j$, $j'$ is chosen from the set of unused ROI indices such that $L[i]=j'$ gives the lowest cost. Legalization often increases cost and therefore does not directly drive minimization, but it converts illegal labelings back to legal ones so that the population of legal assignments can advance.
    \item[V. ]Pooling: Pooling combines $\mathcal{L}$ with the labelings returned from steps I-IV into $\mathcal{L}_{\text{pool}}$ and removes any duplicates.
    \item[VI. ]Culling: Culling is carried out on $\mathcal{L}_{\text{pool}}$ using a modified cost function: $\overline{\text{cost}}(L(\hat{\mathbf{x}})) = \text{cost}(L(\hat{\mathbf{x}})) + k(N-\tilde{N}(L))^2$, where $k$=4 and $\tilde{N}(L)$ is the number of unique $j$ in $L$. The added term does not influence legal assignment costs ($N=\tilde{N}(L)$) but adds a growing penalty for re-used $j$ in $L$ ('illegality') in order to prevent exceedingly non-physical, low-cost labelings from misdirecting the population. With the pool ranked by $\overline{\text{cost}}$, $\mathcal{L}_{\text{new}}$ is built from the set union of: \textit{i}) the 24 best labelings (illegal or legal) and \textit{ii}) the 12 best legal labelings.
    \item[VII. ]Iterating: (Optional) The next cycle is carried out with $\mathcal{L} := \mathcal{L}_{\text{new}}$.
\end{enumerate}

The PM scheme was seeded with $L^{\text{HU}}$ and the naive Bayes labeling $L^{\text{NB}}$ (where $L^{\text{NB}}_i=\argminH_j c_i(\hat{\mathbf{x}}_j)$ minimizes the linear cost sum \Cref{eq:costSum} irrespective of legality) from the corresponding CMV model. Only legal labelings are considered valid output, e.g. for use in \Cref{eq:marginalized} or \Cref{eq:marginalized_truncated}. 

\begin{figure*} [!htb]  %\centering
\includegraphics[width=0.7\textwidth]{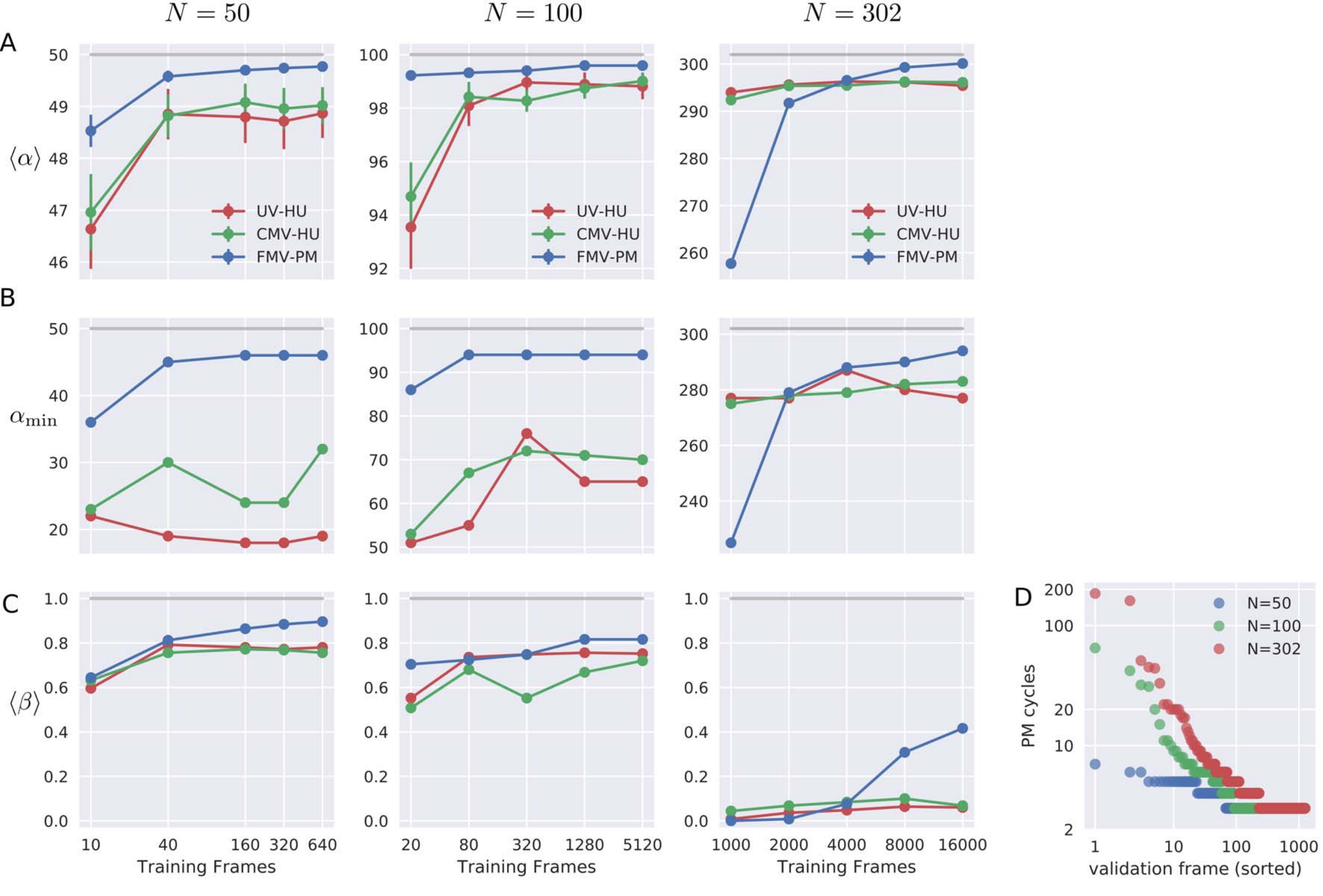}
\caption{\small{(A-C) Atlas guided labeling accuracies. The systems with $N$=50, 100, and 302 cells all had validation sets of 250 frames. (A) The mean labeling accuracy $\left<\alpha\right>$. Error bars are 95\% confidence intervals. (B) The minimum accuracy $\alpha_{\text{min}}$. The y-axis limits differ between (A) and (B) because $\alpha_{\text{min}}$ values have a larger range than $\left<\alpha\right>$ (C) The GT hit fraction $\left<\beta\right>$. Horizontal gray lines in (A)-(C) indicate the theoretical maximum (either $N$ or 1), and the legends in row (A) apply to all other panels in (B)-(C). (D) The number of PM cycles required (a minimum of three cycles were always run) for $\text{cost}(L(\hat{\mathbf{x}})) \le \text{cost}(L^{\text{GT}}(\hat{\mathbf{x}}))$ for frames in the validation set.} }
\label{fig:fig05}
\end{figure*}

\emph{Results.} Here, we compare the GM labelings $L^{\text{GM}}$, found using the different models (UV/CMV/FMV) and labeling schemes (HU/PM). For the UV/CMV models, $L^{\text{HU}}$ is guaranteed to be $L^{\text{GM}}$, whereas the PM search using the FMV model is not guaranteed to find the GM. In the latter case, the best found labelings are considered putative global minima.

To quantify the accuracy of a labeling $L$ relative to $L^{\text{GT}}$ we use two metrics
\begin{equation}
  \begin{aligned}
    \alpha(L)&=&\sum_{i=1}^{N}\delta(L_i-L_i^{\text{GT}}), \\
    \beta(L)&=&\prod_{i=1}^{N}\delta(L_i-L_i^{\text{GT}})
  \end{aligned}
\end{equation}
which are the number of correct labels and a test for an exact match to $L^{\text{GT}}$, respectively. When averaged for a set of validation frames, $\left<\alpha\right>$ is the mean labeling accuracy and $\left< \beta \right>$ is the GT hit fraction.

In Figure \ref{fig:fig05} we plot $\left<\alpha\right>$, $\alpha_{\text{min}}$ (the least accurate labeling in the entire validation set), and $\left<\beta\right>$ as a function of how many frames were used for atlas training. For the $N=50,100$ systems, the FMV model has equal or superior performance to the UV and CMV models. This shows that the correlations between cells' positions, learned by FMV model, carry additional, useful information for cell labeling. The gain in accuracy is most evident for $\alpha_{\text{min}}$, reflecting the fact that the UV and CMV models are susceptible, on occasion, to large assignment errors that cascade through a structure, whereas the FMV model is not. Figure \ref{fig:fig04}C shows one such case where the CMV-HU labeling has $\alpha=$ 81/100. Using the FMV model and PM algorithm, however, the population $\mathcal{L}$ reaches lower cost labelings and eventually finds $L^{\text{GT}}$ after 11 cycles. Despite the high mean accuracy, $\left< \alpha \right>>98\%$, for the $N=100$ system with a well-trained atlas, the fact that $\left< \beta \right> \leq 0.8$ suggests that perfect labeling accuracy remains a challenging goal for an atlas using only position data.

For $N=302$, the FMV model requires a considerable amount of training, >4000 frames, before it outperforms the CMV and UV models on all three accuracy metrics. The high training cost for the FMV model is partly attributable to the need to learn a $906\times906$ covariance matrix for 302 cells in 3D, whereas the $N=100$ 2D system has a $200\times200$ covariance matrix. Moreover, unlike the $N=$ 50 and 100 systems where the cells form a compact cluster, the cells in the $N=302$ system form many smaller clusters that are weakly coupled. In this latter case, it is to be expected that learning the high-dimensional covariance structure requires more training data. If the covariance for real data is similarly high-dimensional, this suggests that a FMV atlas will require a large amount of training data, spanning multiple experiments.

For the FMV results reported above, the PM scheme always found a legal labeling $L$ with $\text{cost}(L(\hat{\mathbf{x}})) \le \text{cost}(L^{\text{GT}}(\hat{\mathbf{x}}))$, which suggests that the combinatorial search is tractable for $N=302$. As Figure \ref{fig:fig05}D shows, this typically required 20 or fewer cycles; however, in a handful of cases (<1\%) with very inaccurate seed labelings, upwards of 200 cycles were required. Each PM cycle for the $N$=302 system required around three minutes on a single cpu core (Intel i9-7940x).

%============================================================================================
\section{Generation of probabilistic output}

We turn to the task of probabilistic output, as in Figure \ref{fig:probatlas}, computed using labelings found during the PM global minimum search or, for the $N=10$ system, computed using all $N!$ labelings. 

\begin{figure*}[!htb] 
    \centering
    \includegraphics[width = 0.95\textwidth]{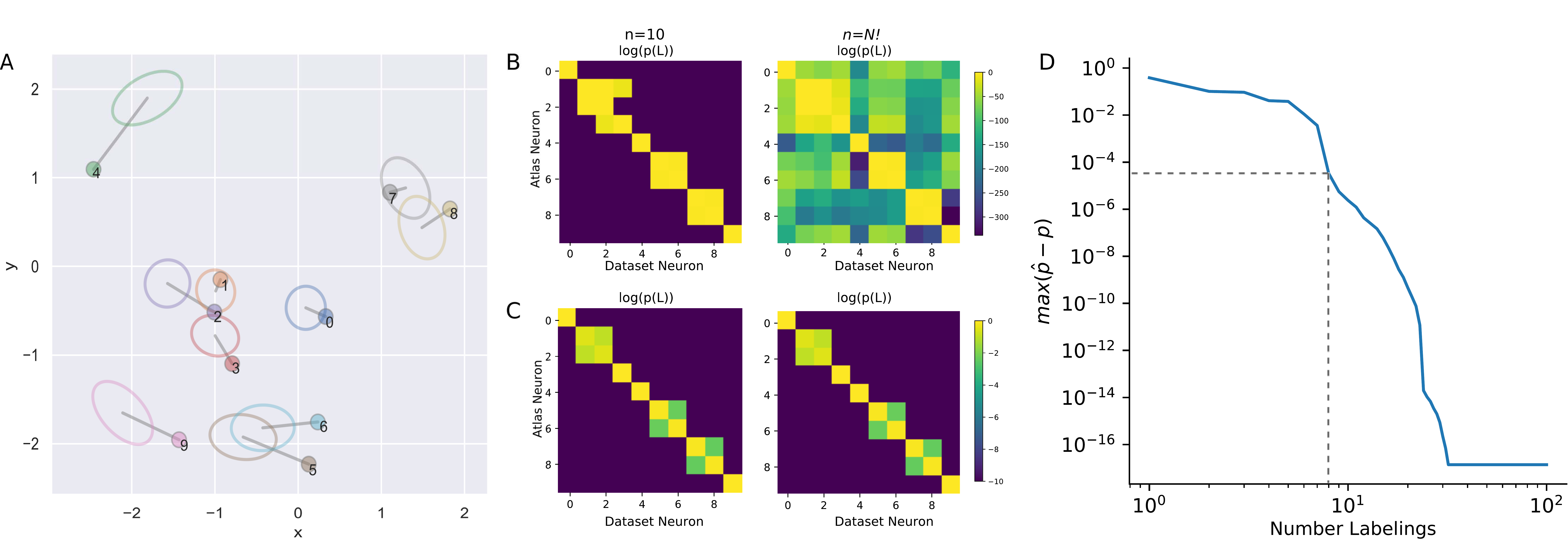}
    \caption{ Evaluation of marginalized probabilities $p_{ij}$ as defined in \Cref{eq:marginalized}  on toy system containing $N=10$ cells. (A) One frame of $N=10$ system. The set of points represent an instance of an incoming dataset, and ellipses represent single cell variances (1$\sigma$ contours). (B) The log marginalized probability $\hat{p}(L(\hat{\mathbf{x}}))$ of an atlas cell (row) being assigned to a dataset cell (column) evaluated on 10 (left) and $N!$ labelings (right). (C) The same log marginalized probability as in (B) thresholded thresholded such that only $p_{ij}>0.005\%$ are reported. Note that there are no clear observable differences between these cases. (D) The maximum absolute error between fully evaluated and truncated marginal probabilities. Evaluating only 8 low-cost labelings (gray dashed line) is sufficient to have a maximum error below 0.01\%. }
    \label{fig:fulleval}
\end{figure*}

In the toy system with 10 point masses, we constructed a matrix of marginal probabilities by fully evaluating $N!$ labelings and comparing them to marginal probabilities estimated from only the 10 lowest cost labelings (Figure \ref{fig:fulleval}). While increasing the number of labelings evaluated does provide a more complete picture of the model output (Figure \ref{fig:fulleval}B), a majority of the $p_{ij}$ have values less than $0.005\%$ and do not significantly improve our output (Figure \ref{fig:fulleval}C). In this illustrative example, sampling as few as 8 low-cost labelings yields a max error of 0.01\%, which increases our confidence in our reporting of truncated marginal probabilities (Figure \ref{fig:fulleval}D). 

\begin{figure*}[!htb] %\centering
\includegraphics[width=0.7\textwidth]{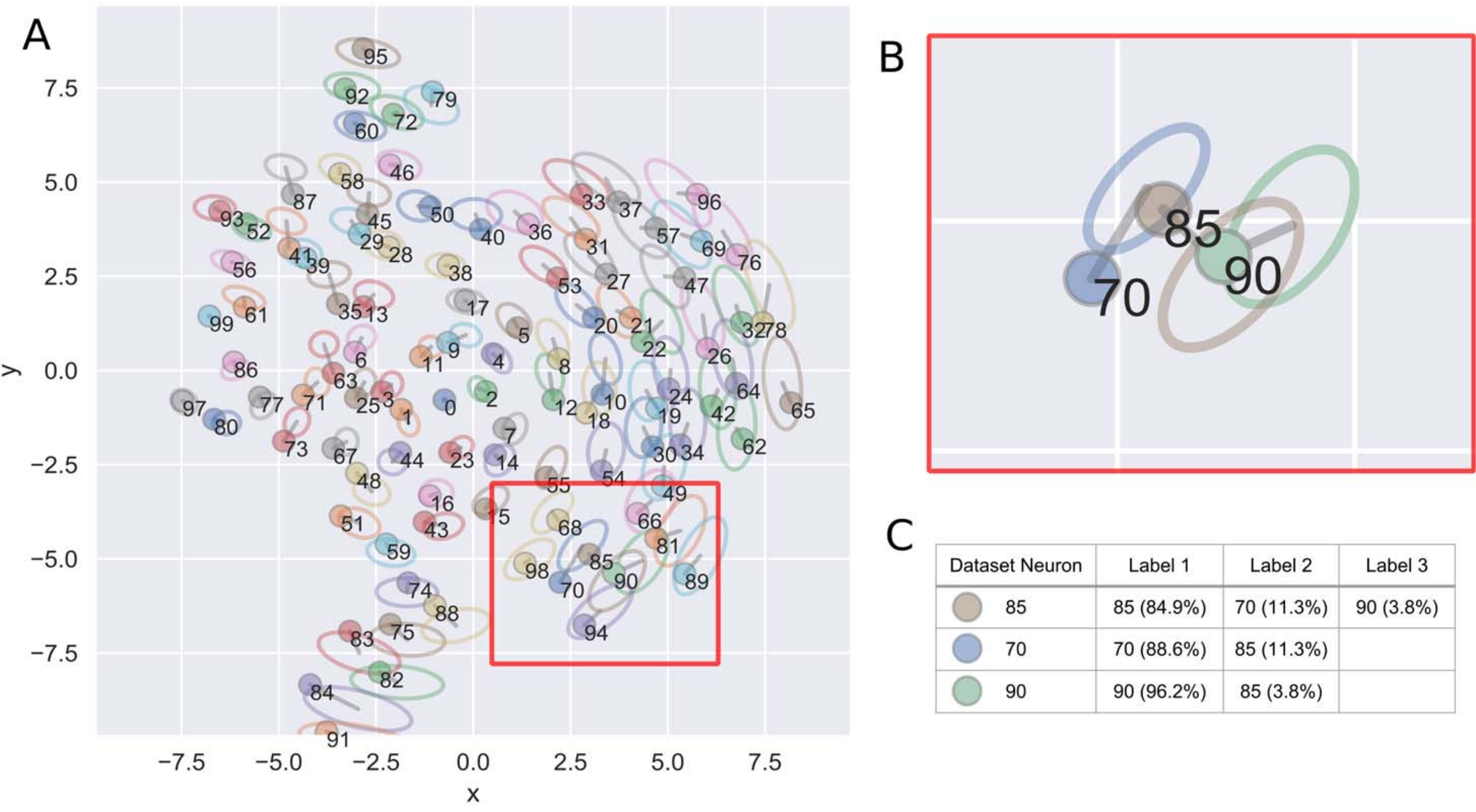}
\caption{\small{ Probabilistic labeling. (A) One frame of the $N=100$ system where the both the NMV-HU and FMV-PM minimum cost labelings are the GT. (B) ROIs (cells) 70, 85, and 90 that have shifted positions relative to their atlas means. In (A) and (B) the formatting is as in Figure \ref{fig:fig04}. (C) The $p_{ij}$ probabilities computed using equation (\ref{eq:marginalized_truncated}) using labelings collected during the PM search. Only labelings contributing $>0.01\%$ were included when computing $p_{ij}$ values. Labels not shown in the table were $>99.99\%$ certain.}}
\label{fig:fig06}
\end{figure*}

Figure \ref{fig:fig06} shows one case for a frame of the $N=100$ system where a probabilistic labeling report was computed using \Cref{eq:marginalized_truncated}. In this case, both the CMV-HU and FMV-PM labelings matched $L^{\text{GT}}$, but a group of neighboring cells (70, 85, 90) have shifted positions relative to their atlas means such that the labeling looks questionable. Without knowing $L^{\text{GT}}$, the table in \Cref{fig:fig06}C captures the uncertainty in the labeling. 

%============================================================================================
\section{Multi-feature atlas}

\begin{figure*}[!htb] %\centering
\includegraphics[width=0.5\paperwidth]{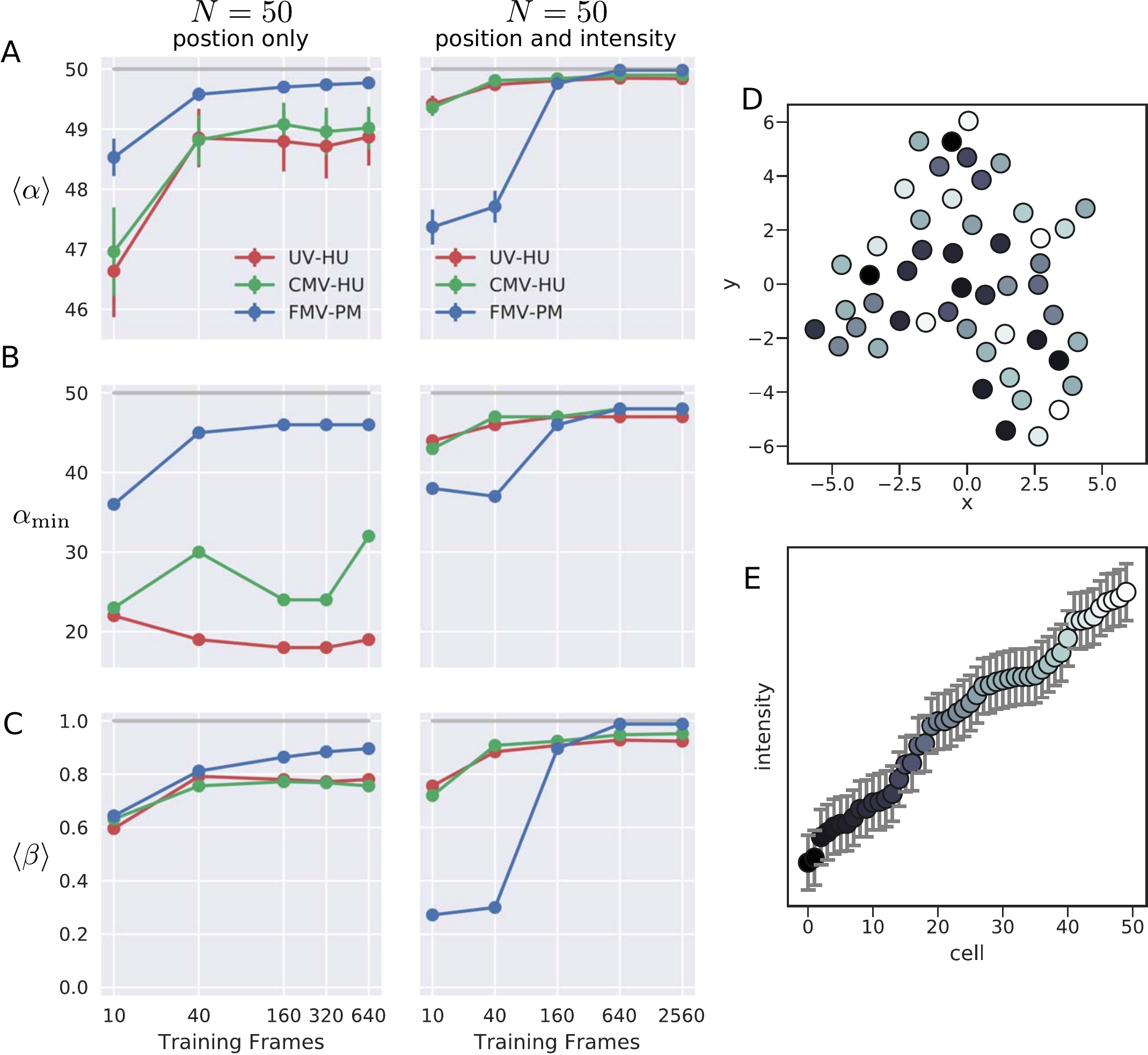}
\caption{\small{Boost in labeling accuracy for a position + fluorescence intensity atlas. (A) Mean labeling accuracy. (B) Minimum labeling accuracy. (C) GT hit fraction. In (A),(B) and (C), the legends and formatting are as in Figure \ref{fig:fig05}. (D) Mean cell positions shaded by mean fluorescence intensity. (E) Cell fluorescence intensities (sorted). Error bars denote two standard deviations.}}
\label{fig:fig08}
\end{figure*}

Lastly, we tested what labeling accuracy might be achievable using an atlas with two feature types, position and cell fluorescence intensity. If cells' fluorescence intensities have a consistent dependence on cell identity across animals, are spatially uncorrelated, and not too noisy, they could help distinguish neighboring cells and improve labeling performance (Figure \ref{fig:fig08}D). For this test, we re-used the $N$=50 system simulations and added a fluorescence intensity to each cell. Intensity was drawn from a uniform distribution across cells with Gaussian noise added to each frame, as shown in Figure \ref{fig:fig08}E. 

In the well-trained limit, all three accuracy metrics $\left< \alpha \right>$, $\alpha_{\text{min}}$, $\left< \beta \right>$ increase with the addition of the intensity feature. For the UV and CMV models, the considerable increase in $\alpha_{\text{min}}$ shows that large misassignment errors are prevented. The GT hit fractions $\left< \beta \right>$ also increase by $\sim10\%$.

The FMV model is inaccurate when only 10-40 training frames are used. We attribute this to the model initially learning spurious correlations between cell positions and fluorescence intensities. With sufficient training, however, these correlations fade and the FMV model outperforms the UV and CMV models. With 640 or more training frames, the FMV model has near perfect labeling accuracy: 247/250 validation frames are perfectly labeled and the other 3/250 have a single erroneous pairwise flip.

This test suggests that an extra feature uncorrelated to other features could substantially increase atlas labeling performance for real data. Multi-feature datasets could also be used to produce labeled datasets to train a position-only atlas for other experimental contexts lacking intensity data. We suspect that the addition of additional fluorescence intensity channels could further increase prediction accuracy. 

%============================================================================================
\section{Discussion and challenges}

We have proposed a machine learning framework to train a probabilistic cell atlas for use in computing probabilistic cell labelings for video or still imaging data. We choose a standard Bayesian parametric probability model, although non-parametric and non-Bayesion approaches could be considered as long as they generate a probabilistic labeling. We used synthetic models of point-masses connected by springs to simulate the biological variability of cells across organisms and experiments, and the collective movements caused by deformations of the surrounding tissue. Using this dataset, we compared different probabilistic atlas models (UV, CMV, FMV) and labeling schemes (HU, PM) for their labeling accuracy. 

We found that a well-trained FMV atlas that models the covariance between cell positions outperforms more simplistic models (UV/CMV) that do not track inter-cell covariance. Standard linear assignment solvers could not be used with the FMV model, due to its non-additive cost function; therefore we developed an algorithm (PM) to carry out the combinatorial labeling search for the FMV atlas.

Using an atlas trained only with cell positions, we found that the single best-guess labeling $L^{\text{GM}}$ was not reliably accurate. However, an atlas that also tracked fluorescence intensity yielded an increase in labeling accuracy, suggesting that utilization of multiple cell features may yield significantly higher performance for real data.

The limited predictive accuracy of a single labeling permutation suggests the need for a probabilistic output that reports label guesses and their uncertainties. We illustrated a means to compute such a probabilistic labeling output. The accuracy of this probabilistic output depends on the extent to which the low-cost labelings (including $L^{\text{GM}}$) are identified and how well the atlas has been trained. For the synthetic $N=10$ case, probabilistic outputs considering the 8 lowest cost labelings were within <0.01\% error of outputs evaluated on all $N!$ labelings. This suggests but does not prove that a reasonably accurate report for the larger systems should not require an intractable amount of combinatorial sampling. For the larger $N=100$ system, the probabilistic assignments as in Figure \ref{fig:fig06} might improve from a more comprehensive search for other low-cost labelings in addition to those collected during the PM search. Ensemble sampling methods such as Wang-Landau sampling (\citealt{Wang:2001}) or simulated tempering (\citealt{Marinari_1992}) could potentially be applied.

In the current scheme, our model is trained on perfectly labeled frames containing the same number of ROIs as atlas labels. We anticipate this not to be the case with data taken from real experiments, which can be expected to have both missing and false-positive ROIs, as well as partial and erroneous labels. Partially supervised learning is an active area of machine learning research.

Furthermore, we assumed that our training data was pre-registered into a common coordinate system. In practice, depending on experimental context, datasets will need to be to pre-processed in a number of ways, each of which could influence labeling outcomes. Achieving high quality registration is an open challenge.

There is a large computational cost of the combinatorial search over labelings and evaluation of multivariate probability models. This cost could be reduced by building a more restricted model tailored to a particular system that leverages locality in physical or feature space. This may allow the use of partial multivariate models that require fewer computations to train and evaluate, as well as efficient means of combinatorial search of labelings. Additionally, approaches based on relaxing the combinatorial search over discrete labelings into a continuous optimization problem (\citealt{paninski:birkhoff,linderman:gumbelsoftmax}) may yield computationally efficient probabilistic labeling algorithms.

The selection of feature vectors and probability distribution family are crucial design decisions of the atlas curator; therefore, we advocate storage of datasets along with the parameterized distribution so that more sophisticated models can be tried later or formal model selection can be performed.

The general probabilistic scheme described here can also be applied to cell typing, rather than cell ID, although the choice of feature vectors, probability distribution families, and labeling algorithms is likely to be different.

%============================================================================================
\section*{Methods}

Each spring mass system consisted of points with cartesian coordinates $\mathbf{x}$ connected by springs. Given initial coordinates $\mathbf{x}^{\text{eq}}$, spring connections were generated to give each point a minimum connectivity, $\text{NN}_{\text{min}}$ to its nearest neighbors. The system's potential energy $U(\mathbf{x})$ is a sum over connected pairs (springs, $S$):
\begin{equation}
    U(\mathbf{x}) = \sum_{(i,j)\in S} \frac{k}{2}(r_{ij}-r_{ij}^{\text{eq}})^2 + a e^{-br_{ij}},
\label{eq:potential}
\end{equation}
where $r_{ij} = ||\mathbf{x}_j - \mathbf{x}_i||$ and $r_{ij}^{\text{eq}}$ is a spring's relaxed distance taken from $\mathbf{x}^{\text{eq}}$ (here $i$ and $j$ index the point-masses within a simulation, and are not to be confused with the previous labeling notation). The exponential term mimics short range inter-cell repulsion and prevents unphysical overlap between connected points. The potential energy parameters for the different toy systems are given in Table II.

Spring-mass system dynamics were integrated using the Langevin equation
\begin{equation}
    m\ddot{\mathbf{x}} = -\nabla U(\mathbf{x}) - \gamma \dot{\mathbf{x}} + \sqrt{2\gamma k_BT}R(t),
\label{eq:Langevin}
\end{equation}
and a Runge-Kutta 4th order integrator with a timestep of $dt$=0.1, where $R(t)$ is a stationary gaussian process, all masses $m$ are set to unity, $k_BT$ is the reduced temperature, and $\gamma$=0.2 provides damping. The system coordinates, snapshotted every 200 integration steps, were used as input for atlas training and validation.

\begin{table}[h]
\caption{Spring-mass simulation parameters}
\label{table:LDparam}
\begin{tabular}{|l|l|l|l|l|l|}
\cline{1-6}
$N$ & $k_BT$ & $k$ & $a$ & $b$ & $\text{NN}_{\text{min}}$         \\
\cline{1-6} 
50   & 1.6 & 10   & 20 & 2   & 7 \\
100  & 1.6 & 10   & 20 & 2   & 7 \\
302  & 1.7 & 0.05 & 20 & 0.2 & 8 \\
\cline{1-6} 
\end{tabular}
\end{table}

All code was implemented in Python and run on a Ubuntu 18.04LTS laptop (Intel i7-7560U) or Windows 10 Home desktop (Intel i9-7940x) and will be available on the focolab.org code repository. The molecular visualization package PyMOL (\citealt{PyMOL}) was used for visualization of spring-mass simulations.

\section*{Acknowledgments}

We thank Sean Escola and Raymond L. Dunn for valuable discussion. This work was supported by the National Institute of General Medical Sciences of the National Institutes of Health under award number R35GM124735 and UCSF grant PBBR-NFR-7028580. The content is solely the responsibility of the authors and does not necessarily represent the official views of the National Institutes of Health.

%\printbibliography %for biblatex
%\section*{References}
%\bibliographystyle{abbrvnat}
\setcitestyle{numbers}

\end{document}